\documentclass[10pt,twocolumn,letterpaper]{article}

\usepackage[pagenumbers]{cvpr} 

\usepackage{amsmath, amssymb, bm}
\usepackage{mathtools}
\usepackage{xspace}        
\usepackage{siunitx}       
\usepackage{glossaries}    
\usepackage{algorithm}
\usepackage{algpseudocode}     
\usepackage{xcolor}
\usepackage{pifont}
\usepackage{arydshln}

\newcommand{\R}{\ensuremath{\mathbb{R}}}
\DeclareMathOperator*{\argmin}{arg\,min}

\newcommand{\maptoR}{\ensuremath{\!:\!\Omega\!\to\!\mathbb{R}}}

\usepackage{graphicx} 
\usepackage{adjustbox}

\newcommand{\boxsym}{\scalebox{0.5}{$\Box$}}

\newcommand{\cm}[1]{\textcolor{teal}{#1}}
\newcommand{\learn}[1]{{\color{red}#1}}

\makeatletter
\renewcommand{\paragraph}{%
    \@startsection{paragraph}{4}%
    {\z@}{-0.5em}{-0.5em}%
    {\normalfont\normalsize\bfseries}%
}
\makeatother

\definecolor{cvprblue}{rgb}{0.21,0.49,0.74}
\usepackage[pagebackref,breaklinks,colorlinks,allcolors=cvprblue]{hyperref}

\usepackage{microtype}
\usepackage{fancyhdr}
\usepackage{url}
\def\confName{CVPR}
\def\confYear{2026}

\title{Solving a Nonlinear Blind Inverse Problem for Tagged MRI \\ with Physics and Deep Generative Priors}

\author{Zhangxing Bian\textsuperscript{1}\hspace{5mm}
 Shuwen Wei\textsuperscript{1}\hspace{5mm} 
 Samuel W. Remedios\textsuperscript{1} \hspace{5mm}
Junyu Chen\textsuperscript{2}  \\
 Aaron Carass\textsuperscript{1}\hspace{5mm} 
 Blake E. Dewey\textsuperscript{2} \hspace{5mm} 
Jerry L. Prince\textsuperscript{1} \\
[2mm]
Johns Hopkins University\textsuperscript{1} \hspace{5mm} 
Johns Hopkins School of Medicine\textsuperscript{2} }

\begin{document}
\twocolumn[{
\maketitle
\begin{center}
\includegraphics[width=\linewidth]{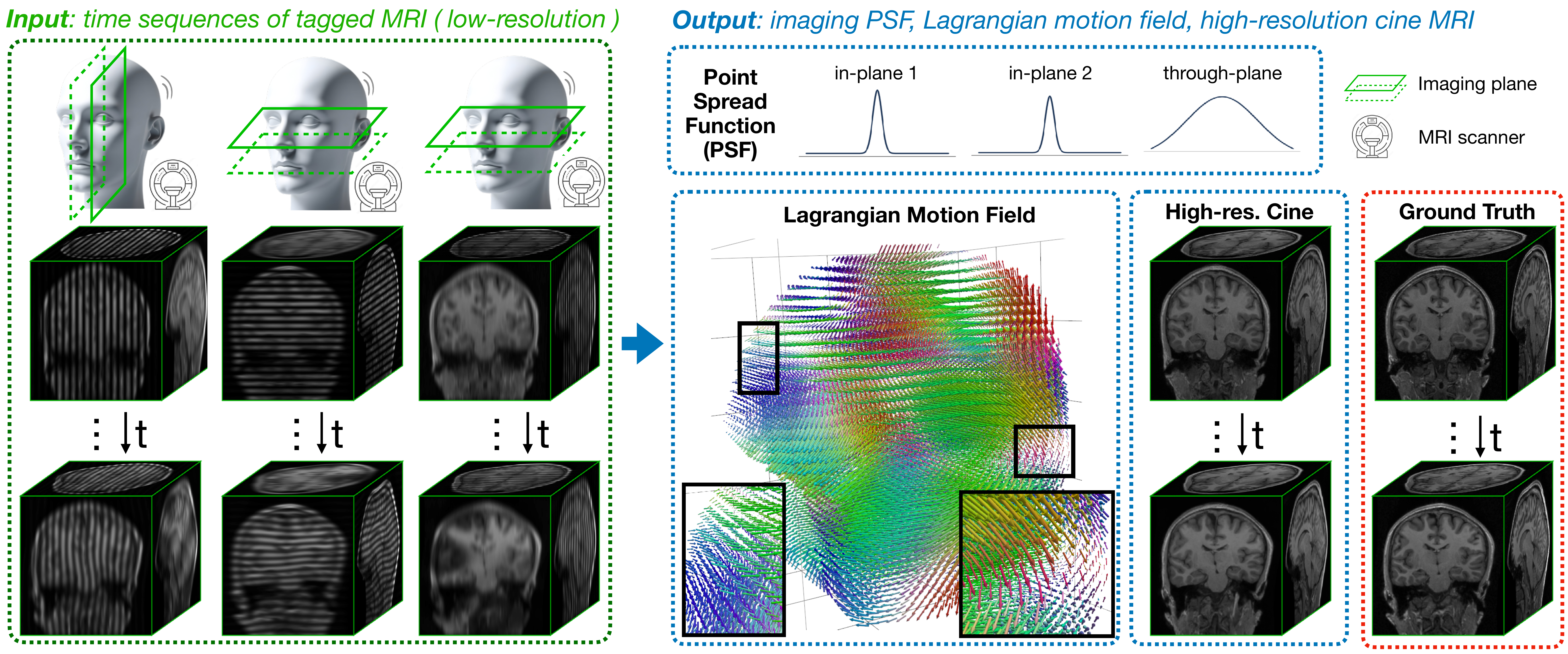}
\end{center}
\vspace{-0.5cm}
\captionsetup{type=figure}
\captionof{figure}{%
We address a nonlinear blind inverse problem: given a time series of 3D tagged MRI volumes, we jointly estimate the imaging point spread functions responsible for image blurring, a sequence of high-resolution cine images, and the underlying motion fields. No additional training data are required.
}\label{fig:teaser}
\vspace{0.3cm}
}]

\begin{abstract}
Tagged MRI enables tracking internal tissue motion non-invasively. It encodes motion by modulating anatomy with periodic tags, which deform along with tissue. However, the entanglement between anatomy, tags and motion poses significant challenges for post-processing. The existence of tags and imaging blur hinders downstream tasks such as segmenting anatomy. Tag fading, due to T1-relaxation, disrupts the brightness constancy assumption for motion tracking. For decades, these challenges have been handled in isolation and sub-optimally. In contrast, we introduce a blind and nonlinear inverse framework for tagged MRI that, for the first time, unifies these tasks: anatomical image recovery, high-resolution cine image synthesis, and motion estimation.
At its core, the synergy of MR physics and generative priors enables us to blindly estimate the unknown forward imaging models and high-resolution underlying anatomy, while simultaneously tracking 3D diffeomorphic Lagrangian motion over time.
Experiments on tagged brain MRI demonstrate that our approach yields high-resolution anatomy images, cine images, and more accurate motion than specialized methods. 
\end{abstract}

\section{Introduction}
Tracking internal tissue motion non-invasively is a central topic in medical imaging. 
Magnetic resonance (MR) tagging encodes \emph{internal} tissue motion by imprinting a temporary periodic pattern of saturated magnetization (i.e.,~tags)~\cite{axel1989mr}. As the tissue moves, the tag lines deform, enabling a visual and quantitative handle on motion and strain. Tagged MRI is widely used for cardiac motion analysis~\cite{axel1989heart, prince1992motion,  ibrahim2011myocardial}, muscle biomechanics~\cite{niitsu1994tongue, moerman2012validation, bian2024drimet}, and brain biomechanics~\cite{knutsen2014improved, bayly2021abe}. However, analyzing tagged MRI is challenging due to artifacts from the tag patterns and imaging process. The tag contrast fades drastically over time (``tag fading'') due to T1 relaxation~\cite{prince2006medical} and progression to steady-state~\cite{bian2024registering}, violating the brightness constancy assumption for optical-flow-based approaches.
Moreover, the spatial resolution is often sacrificed in tagged MR scans compared to standard structural MRI such as T1-weighted (T1w) for imaging speed. Furthermore, due to the existence of tags, most anatomical segmentation methods cannot be directly applied for downstream analysis, which leads to the current practice of acquiring an additional sequence of cine images~\cite{pattynama1993left, bloomer2001cine}. Cine images visualize deforming anatomical structures without tag lines, however, they cost extra acquisition time. These factors make it difficult to accurately recover motion from tagged images and limit its use.

For decades, these challenges have been treated as \emph{separate} tasks in tagged MRI analysis: track motion, synthesize cine images, and super-resolve images. However, this separate treatment often leads to inconsistency and suboptimal use of the data. 
For example, for motion tracking, optical flow~\cite{prince1992motion, dougherty1999validation} and image registration~\cite{ye2021deeptag, ye2023sequencemorph, bian2024drimet} have been adapted for tagged MRI, but they still struggle because tag fading violates the brightness constancy assumption~\cite{bian2024registering}. To address that, Fourier-based approaches~\cite{osman1999cardiac,qian2006extraction,arts2010mapping,wang2013analysis,PVIRA,mella2021harp,bian2024drimet} see tagged images as the amplitude modulation of underlying anatomy with a periodic carrier signal (i.e.,~tags). They extract harmonic peaks to compute and track phase images over time. However, due to the low-frequency nature of the tags, the DC and harmonic spectrum of tagged MRI overlap; tag fading and large deformations will further smear out harmonic peaks and make overlap worse, which violates the assumptions of these Fourier-based methods~\cite{bian2025BRITE}.

Beyond motion tracking, synthesizing tag-free cine MRI from tagged scans reduces the need to acquire a separate cine MRI, simplifying the current clinical workflow.  Recent work~\cite{liu2023attentive, liu2024tagged} treats it as a 2D image-to-image translation. Meanwhile, enhancing the spatial resolution of cine MRI can improve downstream tasks such as segmentation and registration. While super-resolution (SR) has been widely studied in natural~\cite{lu2022transformer,zheng2024smfanet,li2025diffusion} and MR images~\cite{woo2012reconstruction,zhao2019applications,remedios2023self,lei2023decomposition,li2024rethinking}, directly applying frame-independent SR to a cine sequence \emph{risks} introducing temporal inconsistencies and motion-incongruent details---for example, hallucinated edges/structures that vary across frames and break the physical continuity of tissue trajectories.  Although tagged-to-cine synthesis and cine SR have been studied, there has been little work on directly recovering a sequence of \emph{high-resolution 3D} cine from \emph{low-resolution} tagged MRI.

We contend that a unified approach is needed to solve all three tasks \emph{jointly}, as they are fundamentally coupled: (1)~reliable motion tracking requires properly handling tag fading and spectral overlap; (2)~resolving spectral overlap, in turn, requires disentangling the underlying anatomy~(DC spectrum) from the faded tag pattern~(harmonic spectrum); and (3)~producing temporally consistent cine images depends on accurate motion estimates. Moreover, tagged MRI typically trades spatial resolution for imaging speed, yielding low-resolution data. Super-resolving the cine anatomy is beneficial for robust downstream analysis. A unified formulation places these components into a \emph{synergistic} loop, where each strengthens the others.

To this end, we propose InvTag, a \emph{nonlinear} and \emph{blind} inversion framework that takes  \emph{solely} the raw low-resolution tagged MRI data and recovers (1)~a high-resolution anatomy; (2)~a tag-free cine sequence depicting the deforming anatomy; (3)~a bio-mechanically plausible 3D motion field; and (4)~the anisotropic point spread functions of imaging systems. 
The problem is \textit{nonlinear} due to the nonlinear diffeomorphic tissue deformation, and \textit{blind} because key forward-model components (PSF, tag parameters, and time-varying fading) are unknown.

To solve the ill-posedness of the inverse problem, InvTag integrates four ingredients: (1)~a physics-based MRI forward model parameterized by the unknown PSF and tag variables; (2)~a pretrained diffusion prior on widely available high-resolution T1-weighted MRI that regularizes anatomy toward realistic structure; (3)~a physics-informed neural network that estimates smooth, diffeomorphic motion fields; and (4)~a \emph{coordinate descent with diffusion prior}~(CDDP) scheme that alternates diffusion-posterior anatomy sampling with maximum-likelihood updates of unknown parameters. In this framework, the MR physics imposes \emph{hard} constraints (e.g., how blur and tag patterns are formed and fade), while the diffusion prior provides \emph{soft} data-driven constraints (e.g., anatomical plausibility). 
Importantly, InvTag requires no external tagged or cine training data---no paired supervision or finetuning---making it practical in domains where such acquisitions are expensive and scarce.

To our knowledge, InvTag is among the first frameworks that couple MR physics with diffusion generative priors to address a nonlinear, blind inverse problem.
Prior approaches~\cite{chung2022diffusion, zheng2025inversebench-1e1} typically assume linearity or a \emph{known} forward operator, an assumption that is often unrealistic in practice. Our contribution moves beyond this setting by estimating \emph{unknown} imaging parameters jointly with anatomy and motion, thereby extending diffusion-guided inversion to practical blind scenarios.

To summarize, our contributions are: (1)~InvTag, the first unified solution that jointly recovers high-resolution anatomy, cine images, and motion from 3D tagged MRI. (2)~We formulate and solve a practical \emph{nonlinear, blind} inverse problem with MR physics and deep generative prior. (3)~We propose the CDDP scheme that yields stable inversion optimization. (4)~Evaluated on tagged head MRI, InvTag produces high-fidelity super-resolved anatomy, temporally consistent cine synthesis, and improved motion-tracking accuracy compared to specialized methods.

\section{Related Work}
\paragraph{Tagged MRI motion tracking.}
Tagged-MRI motion tracking methods generally operate in either the Fourier domain~\cite{osman1999cardiac,qian2003segmenting,qian2006extraction,huang2008tag,arts2010mapping,wang2013analysis,PVIRA,mella2021harp,bian2024drimet} or image domain~\cite{prince1992motion,dougherty1999validation,ye2021deeptag,ye2023sequencemorph}. 
Fourier-based approaches---such as HARP~\cite{osman1999cardiac,PVIRA,mella2021harp,bian2024drimet,liu2011incompressible}, Gabor filters~\cite{qian2003segmenting,qian2006extraction}, and SinMod~\cite{arts2010mapping,wang2013analysis}---compute motion from the phase of a band-pass-filtered tagged image, assuming distinct harmonic peaks. 
In practice, tag fading, large deformations, and through-plane motion cause spectral overlap~\cite{CSPAMM1993,elsayedreview,bian2024registering}, violating this assumption and leading to tracking errors. 
BRITE~\cite{bian2025BRITE} addresses spectral overlap by explicitly separating the tag pattern and underlying low-resolution anatomy but is limited to 2D and has only been validated on simple anatomies such as a cylindrical phantom. 

\paragraph{Cine synthesis.}
Reconstructing a tag-free cine MRI sequence from tagged data can eliminate the need for a separate cine acquisition. 
HARP-based demodulation~\cite{osman2000imaging,PVIRA} uses the magnitude of the HARP-filtered complex image to approximate a tag-free image but typically yields low spatial resolution. 
More recently, deep learning methods have formulated tagged-to-cine translation as an image-to-image synthesis task. 
Liu et al.~\cite{liu2023attentive,liu2024tagged} trained a Transformer that takes a sequence of 2D tagged images and directly synthesizes the corresponding cine frames. 
However, such data-driven approaches rely on large tagged and cine datasets (either paired or unpaired) for training and may hallucinate inconsistent anatomical structures across time.

\paragraph{Super-resolution.}
Given acquired cine MRI, prior work has sought to enhance its spatial resolution. 
Woo et al.~\cite{woo2012reconstruction} combined three orthogonally oriented 3D volumes with high in-plane resolution to reconstruct a high-resolution cine volume. 
Although numerous super-resolution (SR) methods exist for both natural and MR images~\cite{zhao2019applications,remedios2023self,lei2023decomposition,li2024rethinking,lu2022transformer,zheng2024smfanet,li2025diffusion}, most operate on individual images independently. Directly applying these methods to our task ignores \emph{temporal coherence} across frames and the complementary information provided by \emph{multi-orientation} acquisitions at each time point.

\paragraph{Generative prior for inverse problems.}
Our work relates to the growing use of deep generative models as priors for inverse imaging problems. 
Instead of hand-crafted regularizers, Bora et al.~\cite{bora2017compressed} first demonstrated that a pretrained generative model can serve as a prior for compressed sensing by finding an image within the generator's range that best matches the measurements. 
More recently, plug-and-play diffusion models have become state-of-the-art for many restoration tasks. 
Methods such as DPS~\cite{chung2022diffusion} and many others~\cite{zhu2023denoising,wu2024principled,zhang2025improving,xu2025rethinking} use pretrained diffusion models to solve linear and nonlinear inverse problems~\cite{zheng2025inversebench-1e1} without task-specific retraining. 
A unique aspect of our problem is its simultaneous \emph{nonlinearity} and \emph{blindness}, which remain largely unexplored. 
AmbientGAN~\cite{bora2018ambientgan} learned data distributions of forward operators, and Levac et al.~\cite{levac2025double} extended this idea to jointly learn image and blur-kernel distributions for blind deblurring. 
In contrast, our method tackles a highly nonlinear and multidimensional inverse problem---jointly solving for 3D blurring, multiplicative tag patterns, high-resolution anatomy, and nonlinear motion---all in a blind manner, which has not been previously addressed.

\section{Method}
\label{sec:method}

\subsection{Inverse problem formulation}
\paragraph{Forward model.}
Let $\Omega\!\subset\!\mathbb{R}^3$ be the image domain. For each tag orientation $\boxsym\!\in\!\{\vec{i},\vec{j},\vec{k}\}$ and time $t\!\in\!\{1,\dots,T\}$, the observed tagged image $g_t^{\boxsym}\maptoR$ is modeled as
\begin{align}
    g_t^{\boxsym} &=
    \mathcal{A}^{\boxsym}\!\big(a_t\big) + n_t^{\boxsym}, 
    \label{eq:abstract-forward-1} \\
    &= h_\gamma^{\boxsym} * (q_t^{\boxsym} \cdot a_t) + n_t^{\boxsym},
    \label{eq:abstract-forward-2}
\end{align}
where $a_t\maptoR$ is the underlying deformed anatomy at time $t$, $q_t^{\boxsym}\maptoR$ is the faded and deformed tag pattern  at time $t$, $h_\gamma^{\boxsym}$ is the point-spread function (PSF) for orientation $\boxsym$ (parameterized by $\gamma$), $*$ denotes three-dimensional convolution, $\cdot$ denotes scalar multiplication, and $n_t^{\boxsym}\maptoR$ is noise. The spatial variable $\bm{r} \in \Omega$ is omitted for brevity. Each orientation vector ${\boxsym}$ points \emph{along} the tag propagation direction, i.e., perpendicular to the tag lines. 

We represent motion as a diffeomorphism $\phi_t:\Omega\!\to\!\Omega$ that maps coordinates at time $t$ \emph{to} the \emph{undeformed} reference frame. The corresponding image warp (pullback) operator is defined as
\begin{equation}
    \big(\phi^*u\big)(\bm{r}) = u\big(\phi(\bm{r})\big)\,.
\end{equation}
Using this operator, the deformed anatomy and tag pattern at time $t$ can be written as
\begin{equation}
    a_t \;=\; \phi_t^*\;a,
    \qquad
    q_t^{\boxsym} \;=\; \phi_t^*\;f_t(q^{\boxsym}),
    \label{eq:kinematics-photometric}
\end{equation}
where $a$ denotes the \emph{undeformed} anatomy at the reference frame, $q^{\boxsym}$ is the \emph{undeformed} base tag pattern for orientation $\boxsym$ at the reference frame, and $f_t:\mathbb{R}\!\to\!\mathbb{R}$ models tag fading at time $t$.
Plugging~\eqref{eq:kinematics-photometric} in \eqref{eq:abstract-forward-2} and using the identity ($\phi_t^*\;u)\cdot (\phi_t^*\;v)=\phi_t^*\; (u\cdot v)$, we get
\begin{equation}
    g_t^{\boxsym}
    \;=\;
    h_\gamma^{\boxsym} * \,\phi_t^*\;\left[a \cdot f_t(q^{\boxsym})\right] \;+\; n_t^{\boxsym}.
    \label{eq:forward}
\end{equation}

\paragraph{Inverse problem.}
Given multi–time and multi–orientation measurements $\{g_t^{\boxsym}\}$, the goal is to jointly estimate
\[
a, \quad \{q^{\boxsym}\}, \quad \{h_\gamma^{\boxsym}\}, \quad \{f_t\}_{t=0}^T, \quad \{\phi_t\}_{t=1}^T
\]
such that \eqref{eq:forward} is satisfied for all $(t,\boxsym)$. 
This constitutes a \emph{nonlinear blind} inverse problem: nonlinearity arises from the deformable spatial transformation, while blindness results from unknown imaging parameters and the fading process.
Instead of directly estimating deformed anatomy $a_t$ at every time as in~\eqref{eq:abstract-forward-2}, the formulation~\eqref{eq:forward}  
uses explicit deformation $\phi_t$ to link all frames to a common anatomy $a$, enforcing \emph{temporal anatomical consistency} by construction: anatomy varies geometrically only through $\phi_t$. 

\subsection{Parametrization and assumptions}
To make the system identifiable and numerically solvable, we impose a set of MR physics-informed simplifications on each component of the model.

\paragraph{Anisotropic Imaging PSF.}
We model $h_\gamma^{\boxsym}$ as an \emph{anisotropic} three-dimensional Gaussian with standard deviations 
$\gamma \coloneqq (\gamma{_\perp},\gamma_{\parallel},\gamma_{\mathrm{thru}})$. 
For each acquisition of tag orientation $\boxsym\,$, $\gamma_{\perp}$ and $\gamma_{\parallel}$ control the in-plane blur orthogonal and parallel to the tag lines, respectively, and $\gamma_{\mathrm{thru}}$ controls through-plane blur. The kernel is normalized to preserve mean intensity (unit DC gain) after convolution with an image.
We assume this Gaussian approximation captures the dominant blur effects in routine MRI acquisitions; extension to more complex PSF models (e.g., multi-slice cross-talk) is left for future work.

\paragraph{Base tag patterns.}
The undeformed base patterns $q^{\boxsym}$ follow standard SPAMM physics~\cite{axel1989mr} and are parameterized by sinusoidal functions.
Thus for tag orientation $\boxsym$, we write
\begin{equation}
q_\alpha^{\boxsym} \coloneqq  p_0^{\boxsym}(\bm{r};\alpha)
\;=\;
\alpha_1 \cos\!\left(
2\pi\,\langle {\boxsym}, \bm{r} \rangle / \alpha_2 \;+\; \alpha_3
\right) 
 + \alpha_4 
\label{eq:spamm-11}
\end{equation}
where $\bm{r}\!\in\!\Omega$ is the spatial coordinate variable, and $\langle \cdot, \cdot \rangle$ denotes the inner product. 
The $\alpha \coloneqq \{\alpha_1,\alpha_2,\alpha_3,\alpha_4\} \in \R^4$, where $\alpha_1$ (amplitude), $\alpha_2$ (tag spacing), $\alpha_3$ (phase), and $\alpha_4$ (DC offset) parameterize the base tag pattern. The parameters $\alpha$ are shared across all tag orientations $\boxsym$.
Although nominal tag spacing can be obtained from DICOM metadata (when available), the effective values often deviate due to MR acquisition artifacts.
Thus, we treat the $\alpha$'s as unknown.

\paragraph{Tag fading.}
Fading is governed by the magnetic resonance properties of tissue and imaging protocols, which is assumed to be \emph{unknown} in our formulation. The fading process can be captured by an affine model~\cite{bian2024registering} applied to the base tag:
\begin{equation}
     f_{\beta_t}(q_\alpha^{\boxsym}) \coloneqq f_t(q^{\boxsym}; \beta_t) \;=\; \big|\beta_{1,t} \cdot q^{\boxsym} + \beta_{2,t}\big| ,
    \label{eq:fading}
\end{equation}
with $\beta_t \coloneqq (\beta_{1,t},\beta_{2,t})\in\mathbb{R}^2$ denoting, respectively, amplitude and DC shift at time $t$, shared across all $\boxsym$. The notation $|\cdot|$ denotes absolute value.

\paragraph{Motion.}
We represent the $\phi_t$ as a neural field
$\phi_t(\cdot;\theta_t):\Omega\!\to\!\Omega$, where $\theta_t$ are the weights of a physics-informed neural network~(PINN)~\cite{raissi2019physics}.
The network takes coordinates $\bm{r}$ and outputs a stationary velocity field $\bm{v}_t(\bm{r};\theta_t)$, and 
$\phi_t(\bm{r};\theta_t)$ is computed by an exponential mapping of $\bm{v}_t$: \begin{align}
    \phi_{\theta_t} \coloneqq  \phi_t(\bm{r};\theta_t) = \exp{\{\bm{v}_t(\bm{r};\theta_t)\}} \,.
\end{align} 
The $\exp$ is implemented with fast scaling-and-squaring~\cite{arsigny2006log} integration, 
promoting a diffeomorphic deformation field.

\paragraph{Parametrized forward model.}
From the previous, the \emph{parametrized} forward model for $(t, {\boxsym})$ can be written as 
\begin{equation}
    \mathcal{A}_{t}^{\boxsym}\!\big(a;\alpha, \gamma,\beta_t,\theta_t\big)
\;=\;
h_\gamma^{\boxsym} * \phi_{\theta_t}^*\!\left[\, a \cdot f_{\beta_t}\!\big(q_\alpha^{\boxsym}\big) \,\right]\,.
\label{eq:parametrized-forward}
\end{equation}

\subsection{Coordinate descent with diffusion prior}
Due to the blindness and ill-posedness of the inverse problem, we solve it through a \emph{coordinate descent} strategy. 
The key idea is to iteratively update (A) the underlying anatomy $a$ under a diffusion prior while fixing the forward-model parameters, and (B) the forward-model parameters while fixing $a$. 
Empirically, this alternating scheme yields better numerical stability and convergence robustness than joint optimization over all variables.

\paragraph{(A) Estimate anatomy via diffusion posterior sampling.}
To search for a realistic anatomy $a$ that explains the observations, we impose a diffusion prior and treat the forward model as fixed. For clarity, we will continue to express all variables as continuous functions on $\Omega$, although in practice all computations operate on a discretized voxel grid.

Diffusion models define the generative process as the {\em reverse} of the noising process.
Given observations~$\{g_t^{\boxsym}\}$, we draw samples from the \emph{posterior} $p(a|\{g_t^{\boxsym}\})$ via the reverse-time stochastic differential equation~(SDE)~\cite{song2020score,anderson1982reverse}
\begin{equation}
\small
    da_{\tau}
    =
    - \eta_\tau \left[a_{\tau}/2
    + \nabla_{a_{\tau}} \log p_{\tau}(a_{\tau}|\{g_t^{\boxsym}\})\right]d\tau
    + \sqrt{\eta_\tau}\,d\bar{w},
    \label{eq:reverse-sde-functional}
\end{equation}
where $\tau$ denotes the diffusion time running backward from $1$ to $0$, and $\eta_\tau: {\R \rightarrow \R} > 0$ is the noise schedule of the diffusion process~\cite{ho2020denoising}. $d\bar{w}$ is the standard Wiener process running backward. Here, the subscript in $a_{\tau}$ indexes diffusion time $\tau$---not the physical time $t$.
Using Bayes’ rule,
\begin{equation}
\begin{split}
    \nabla_{a_{\tau}} \log p_{\tau}(a_{\tau}|\{g_t^{\boxsym}\})
    &= \nabla_{a_{\tau}}\log p_{\tau}(a_{\tau}) \\
    &\quad+ \nabla_{a_{\tau}}\log p_{\tau}(\{g_t^{\boxsym}\}|a_{\tau})\, ,
\end{split}
\label{eq:bayes-two_terms}
\end{equation}
where the first term represents the diffusion prior and can be directly computed by a pre-trained score function $s_{\vartheta}(a_{\tau},\tau)$, and the second term enforces data consistency.

The second term in (\ref{eq:bayes-two_terms}) is approximated by DPS~\cite{chung2022diffusion} as
\begin{equation}
\nabla_{a_{\tau}}\log p(\{g_t^{\boxsym}\}|a_{\tau})
\simeq
\nabla_{a_{\tau}}\log p(\{g_t^{\boxsym}\}|\hat{a}_0),
\end{equation}
where $\hat{a}_0$ is an estimate of the posterior mean of $p(a_0|a_{\tau})$ obtained by Tweedie’s formula~\cite{robbins1992empirical, efron2011tweedie,kim2021noisescore}:
\begin{equation}
\hat{a}_0
\simeq
\frac{1}{\sqrt{\bar{\kappa}_\tau}}
\big(a_{\tau} + [1 - \bar{\kappa}_\tau]\,s_{\vartheta}(a_{\tau}, \tau)\big)\,,
\label{eq:hat-a0}
\end{equation}
where $i$ indexes discrete diffusion steps, $\kappa_i \coloneqq 1 - \eta_i$, and $\bar\kappa_i \coloneqq \prod_{j=1}^i \kappa_j$ following~\cite{ho2020denoising}.

Assuming i.i.d.\ Gaussian noise $n_t^{\boxsym}\sim\mathcal{N}(0,\sigma^2)$ in the forward model~\eqref{eq:forward}, the likelihood becomes
\begin{equation}
    p(\{g_t^{\boxsym}\}|a)
\propto
\exp\!\left[-\tfrac{1}{2\sigma^2} \mathcal{L}_\mathrm{rec}(a) \right]\,,
\end{equation}
where 
\begin{equation}
    \mathcal{L}_\mathrm{rec}(a) = 
    \sum_{t} \mathcal{L}_\mathrm{rec}^t(a)
    =
    \sum_{t}\sum_{\boxsym}\!\big\|g_t^{\boxsym} - \mathcal{A}_t^{\boxsym}(a)\big\|_2^2 \, .
\end{equation}
Differentiating with respect to $a_{\tau}$ yields,
\begin{equation}
\nabla_{a_{\tau}}\log p(\{g_t^{\boxsym}\}|a_{\tau})
\simeq
-\rho\,\nabla_{a_{\tau}}
\mathcal{L}_\mathrm{rec}\big(\hat{a}_0(a_{\tau})\big) \, ,
\label{eq:likelihood-grad}
\end{equation}
where we explicitly write $\hat{a}_0:=\hat{a}_0(a_{\tau})$ to emphasize that $\hat{a}_0$ is a function of $a_{\tau}$. 
Consequently, the gradient $\nabla_{a_{\tau}}$ amounts to backpropagation through the network.
Step size $\rho:=1/\sigma^2 \approx \rho^*/\mathcal{L}_\mathrm{rec}$ where $\rho^*$ is a hyper-parameter.
By \eqref{eq:reverse-sde-functional}, \eqref{eq:bayes-two_terms}, and \eqref{eq:likelihood-grad}, we have
\begin{equation}
\begin{split}
da_\tau =& - \eta_\tau \Bigl[\tfrac{1}{2} a_\tau
    + s_{\vartheta}(a_\tau,\tau) \\
    &- \rho\nabla_{a_\tau}\mathcal{L}_\mathrm{rec}\bigl(\hat a_0(a_\tau)\bigr) \Bigr] d\tau + \sqrt{\eta_\tau} d\bar w .
\end{split}
\label{eq:dps-functional}
\end{equation}
which describes the full reverse diffusion process: the term $s_{\vartheta}$ \emph{pulls} the sample toward the ``less noisy'' anatomy manifold, while the term $\nabla_{a_{\tau}}
\mathcal{L}_\mathrm{rec}$ \emph{pushes} it toward data fidelity. Discretizing this reverse SDE and sweeping $\tau$ from $1$ to $0$ yields the updated anatomy $a$.

\paragraph{(B)~Estimate forward model via maximum-likelihood.}
With $a$ fixed, the maximum-likelihood estimates of  $(\alpha,\gamma,{\beta_t},{\theta_t})$ are obtained by minimizing the data residual:
\begin{equation}
\argmin_{\alpha,\gamma,\{\beta_t\},\{\theta_t\}}
\;\mathcal{L}_{\mathrm{rec}}\!\big(a;\alpha,\gamma,\{\beta_t\},\{\theta_t\}\big).
\label{eq:ml-nuisance}
\end{equation}
For the low-dimensional parameters $(\alpha,\gamma,\{\beta_t\})$, we use a bounded population-based optimizer (e.g., differential evolution~\cite{storn1997differential}), which performs robustly in highly nonconvex landscapes and is less sensitive to initialization. Empirically, it converges faster and finds better optima than standard gradient-based methods such as gradient descent or Adam~\cite{kingma2014adam}.
For the high-dimensional motion parameters ${\theta_t}$ (a PINN), we adopt the Adam optimizer for cost-efficiency.

\paragraph{Coordinate descent with diffusion prior~(CDDP).}
Putting the components together, Algorithm~\ref{alg:alt} performs CDDP to solve our nonlinear and blind inverse problem. We initialize the reference anatomy $a$ by averaging the three orientations at $t=1$ and applying a low-pass filter. 
The undeformed base tag is initialized to a constant image of ones, the motion is set to identity mapping ($\phi_{\theta_1}\!=\!\mathrm{Id}$), the PSF to a Dirac kernel ($h_\gamma\!=\!\delta$) with no blur, and the fading function to the identity mapping ($f_{\beta_t}\!=\!\mathrm{Id}$). 
\textbf{At the first timeframe} $t{=}1$, we run $L$ loops over the PSF, tags, anatomy, and motion (lines 5-8)  to estimate $(a^\star,\alpha^\star,\gamma^\star)$, and keep these fixed (line 10) for all later frames ($t{>}1$).
This enforces temporal anatomical consistency and avoids slow drift when tag contrast fades over time. 
The anatomy update uses a pretrained diffusion score $s_\vartheta$ trained on over 80,000 \SI{1}{mm} isotropic T1-weighted 3D head volumes~\cite{remedios2025diffusion}.
At each round~$\ell$, the reverse process~\eqref{eq:dps-functional} samples an updated $a$. During sampling, the pretrained diffusion weights are frozen and 256 DDIM~\cite{song2020denoising} steps are used.
For each \textbf{subsequent frame} $t{>}1$, we alternate $L$ times between fading and motion while keeping $(a^\star,\alpha^\star,\gamma^\star)$ fixed. 
We use $L=4$. 
Motion is initialized from the previous time state ($\theta_t\!\leftarrow\!\theta_{t-1}$), which enables \emph{large} Lagrangian deformation tracking and avoids periodic tag matching ambiguities~\cite{liu2010shortest-f52, bian2023momentamorph}.

Low-dimensional parameters $(\gamma,\alpha,\beta_t)$ are optimized with bounded differential evolution; bounds are chosen to cover typical values encountered in routine MRI.
The bounds for the anisotropic PSF sigmas are $\gamma_{\perp},\gamma_{\parallel},\gamma_{\mathrm{thru}}\in[0,4]$ voxels; the tag parameters use amplitude $\alpha_1\in[0.5,1.]$, spacing $\alpha_2\in[0.9,1.1]\times(\text{nominal spacing})$, phase $\alpha_3\in[-\pi,\pi]$, and DC level $\alpha_4\in[0,0.5]$; the fading parameters obey $\beta_{1,t}, \beta_{2,t} \in[0,1]$.
These bounds are broad enough to capture typical in-plane/through-plane blur, tag spacing deviations from nominal values, and tag fading encountered in routine MRI acquisitions. The PINN motion parameters are optimized with Adam.

\begin{algorithm}[t]
\caption{CDDP for nonlinear blind inverse problem}
\label{alg:alt}
\begin{algorithmic}[1]
\State \textbf{Input:} measurements $\{g_t^{\boxsym}\}$ for $\boxsym\in\{\vec{i},\vec{j},\vec{k}\}$ and $t=1{:}T$; pretrained diffusion score $s_{\vartheta}$
\State \textbf{Output:} anatomy $a$, tag parameters $\alpha$, PSF $\gamma$, fading $\{\beta_t\}$, motion $\{\phi_t\}$
\State \textbf{Initialization:} $a \gets \mathtt{lowpass}\!\left(\mathtt{avg} (g_1^{\boxsym})\right)$;
$q_{\alpha} \gets \bm{1}$;
$\phi_{\theta_1} \gets \mathtt{Id}$;
$h_{\gamma} \gets \delta$;
$f_{\beta_t} \gets \mathtt{Id}$
\medskip
\State \textbf{if} $t{=}1$ \textbf{then for} $\ell=1{:}L$ \textbf{do}
\State \hspace{\algorithmicindent} PSF: $\learn{\gamma} \gets \min_{\gamma}
        \mathcal{L}_{\mathrm{rec}}^t \big(a;\alpha,\learn{\gamma},\beta_t,\theta_t\big)$
\State \hspace{\algorithmicindent} \text{Tags:} $\learn{\alpha} \gets \min_{\alpha}
        \mathcal{L}_{\mathrm{rec}}^t \big(a;\learn{\alpha},\gamma,\beta_t,\theta_t\big)$
\State \hspace{\algorithmicindent} \text{Anatomy:} 
        $\learn{a} \gets \mathtt{sample} \big(\alpha,\gamma,\beta_t,\theta_t, s_{\vartheta} \big)$ as in~\eqref{eq:dps-functional}
\State \hspace{\algorithmicindent} \text{Motion:} 
        $\learn{\theta_t} \gets  \min_{\theta_t}
        \mathcal{L}_{\mathrm{rec}}^t \big(a;\alpha,\gamma,\beta_t,\learn{\theta_t}\big)$
\State \textbf{end for}
\State Set $a^\star\gets a$, $\alpha^\star\gets \alpha$, $\gamma^\star\gets \gamma$
\For{$t=2$ to $T$}
    \State $\theta_t \gets \theta_{t-1}$ \cm{\# init from previous motion state}
    \For{$\ell=1$ to $L$}
        \State \text{Fading:} 
        $\learn{\beta_t} \gets \min_{\beta_t}\ 
        \mathcal{L}_{\mathrm{rec}}^t\!\big(a^\star;\alpha^\star,\gamma^\star,\learn{\beta_t},\theta_t\big)$
        \State \text{Motion:} $\learn{\theta_t} \gets  \min_{\theta_t} \mathcal{L}_{\mathrm{rec}}^t\!\big(a^\star;\alpha^\star,\gamma^\star,\beta_t,\learn{\theta_t}\big)$
    \EndFor
    \State Set $\beta_t^\star\gets \beta_t$, $\theta_t^\star\gets \theta_t$
\EndFor

\State \textbf{Return:} $a^\star$, $\alpha^\star$, $\gamma^\star$, $\{\beta_t^\star\}$, and $\{\phi_{\theta_t}^\star\}$
\end{algorithmic}
\end{algorithm}

\begin{figure*}
    \centering
    \includegraphics[width=1\linewidth]{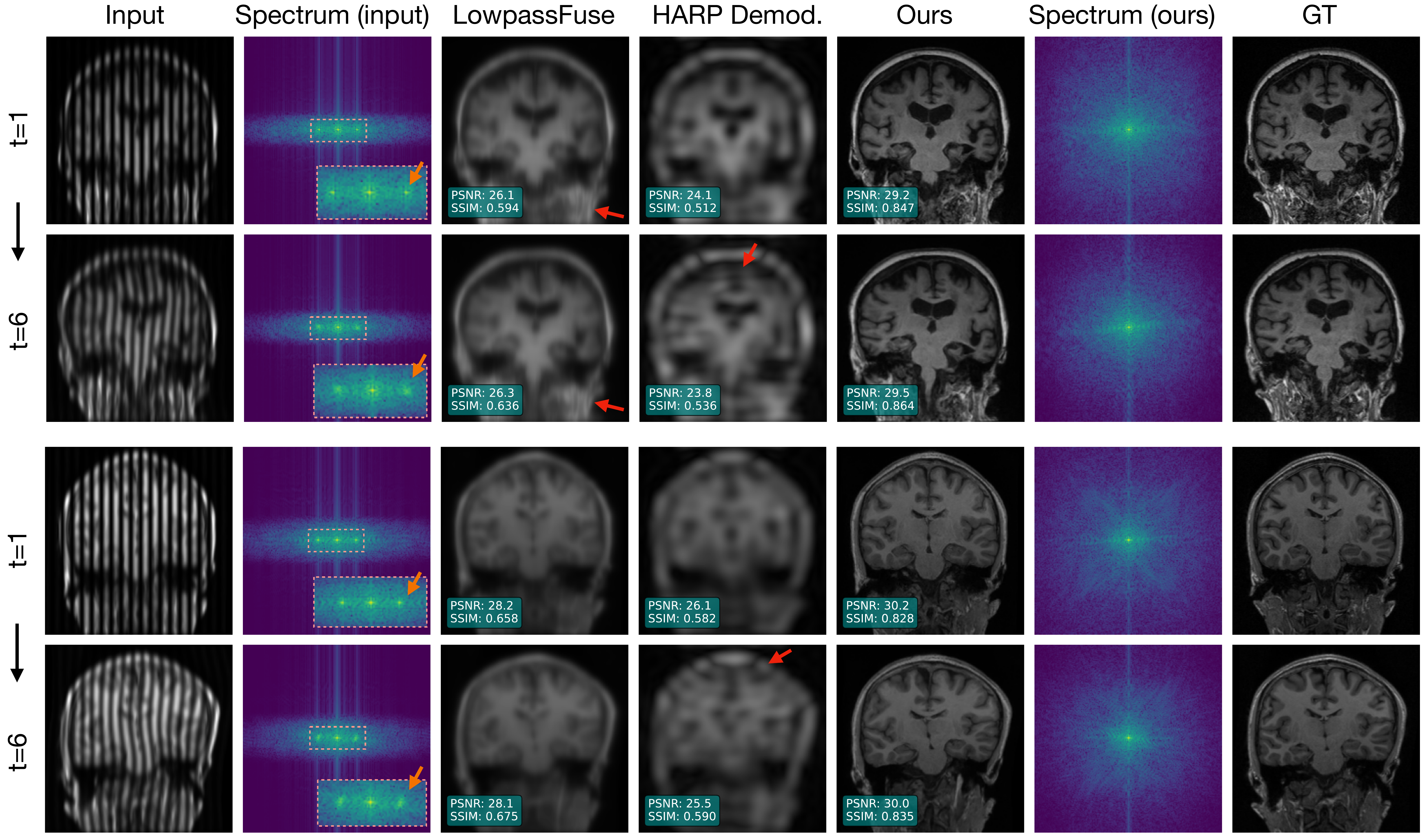}
    \caption{Qualitative comparison of tag-to-cine synthesis. For two subjects, we show early ($t{=}1$) and late ($t{=}6$) frames. Left-most column shows input tagged MRI (other views and orientations are omitted) while right-most shows ground-truth cine. Fourier spectrum of input and ours are shown. Orange arrows indicate the faded and smeared harmonic peaks. Red arrows indicate the aliasing artifact.}
    \label{fig:cine_compare}
\end{figure*}

\paragraph{Implementation Details}
Volumes are center-cropped and/or padded to $192 \times 224 \times 192$ voxels and normalized to $[0, 1]$.
The DDPM uses a five-level 3D U-Net with 256 DDIM sampling steps, pre-trained on over 80{,}000 \SI{1}{mm} isotropic T1w volumes~\cite{remedios2025diffusion}.
For a 6-frame 3D sequence, our pipeline runs \SI{1.2}{h} per frame on a single NVIDIA A40 GPU with a peak memory of \SI{38}{GB}.
Full implementation details are in Appendix~\ref{sec:impl-details}.

\section{Experiment Results}
Our method produces two main outputs: a sequence of high-resolution cine MRIs and motion fields, which we evaluate on the task of tag-to-cine synthesis and motion estimation.

\subsection{Experimental Setup}

\paragraph{Dataset \& Operators.}
We randomly selected 20 unique T1w subjects from each of AIBL\footnote{\scriptsize\url{aibl.org.au/}} and Sleep\footnote{\scriptsize\url{openneuro.org/datasets/ds003826/versions/3.0.1}} for testing. MRIs in Sleep are defaced. 
Neither dataset is included in the DDPM training data; both are reserved exclusively for testing. 
Tag formation and fading follow the MR physics model~\cite{bian2024registering}, and images are blurred with 3D Gaussian kernels. 
For each subject, we independently generate a diffeomorphic deformation by integrating a randomly generated divergence-free velocity field (via curl of vector potentials) using a 4th-order Runge-Kutta scheme, yielding realistic, incompressible motion; the resulting fields are then used to warp the tagged images and create tagged MRI sequences of length $6$. 
To cover imaging variability, we evaluated four settings: three blur configurations \((\gamma_\perp, \gamma_\parallel, \gamma_{\mathrm{thru}})\)---$(1,1,1)$, $(0.4,0.4,4)$, $(0.4,1,3)$---without additive noise (covering isotropic and anisotropic blur), and $(0.4,1,3)$ with Gaussian noise (zero mean, standard deviation $0.01$ of the image dynamic range). 
There are $(20+20) \times 4 = 160$ unique test cases in total. 

\paragraph{Metrics.}

Since we have ground truth (GT) images from MR tagging simulations, we can use reference-based image quality metrics to evaluate tagged-to-cine synthesis and super-resolution performance. We normalize image intensity to the same dynamic range and use peak signal-to-noise ratio (PSNR) and structural similarity index measure (SSIM) in scikit-image implementations~\cite{van2014scikit}.
For motion accuracy, we report the mean end-point error~(EPE), $\mathtt{EPE}(\bm{r}) = \|\mathbf{u}_\mathrm{gt}(\bm{r}) - \mathbf{u}_\mathrm{est}(\bm{r})\|_2$, the displacement error magnitude, and its 95th percentile (EPE@95) to summarize tail errors. We also report the percentage of voxels with negative Jacobian determinants to assess violations of diffeomorphism (i.e., tissue foldings).

\begin{table}
\centering
\scriptsize
\begin{tabular}{lccc}
\toprule
\textbf{Methods} & \textbf{Timeframe} & \textbf{PSNR}$\uparrow$ & \textbf{SSIM}$\uparrow$ \\
\midrule
LowpassFuse   & $t{=}1$ & 26.43 $\pm$ 1.40 & 0.62 $\pm$ 0.06 \\
   & $t{=}6$ & 26.68 $\pm$ 1.39 & 0.66 $\pm$ 0.05 \\
   \cdashline{1-4}\noalign{\vskip 0.6ex}
HARP Demod. & $t{=}1$ & 24.28 $\pm$ 1.26 & 0.52 $\pm$ 0.04 \\
 & $t{=}6$ & 23.93 $\pm$ 1.17 & 0.54 $\pm$ 0.03 \\
 \cdashline{1-4}\noalign{\vskip 0.6ex}
Ours     & $t{=}1$ & \textbf{28.38} $\pm$ \textbf{1.41} & \textbf{0.83} $\pm$ \textbf{0.04} \\
     & $t{=}6$ & \textbf{28.41} $\pm$ \textbf{1.36} & \textbf{0.84} $\pm$ \textbf{0.03} \\
\bottomrule
\end{tabular} \vspace{-1mm}
\caption{Quantitative comparison of tagged-to-cine synthesis across methods and timeframes, reported as mean $\pm$ std.}
\label{tab:synthesis_metrics}
\end{table}

\begin{table}[t]
\centering
\scriptsize
\adjustbox{width=0.49\linewidth}{
\begin{tabular}{lcc}
\toprule
\textbf{blur, noise} & \textbf{PSNR}$\uparrow$ & \textbf{SSIM}$\uparrow$ \\
\midrule
$(1, 1, 1)$, \ding{55}         & 29.15 & 0.85 \\
$(0.4, 0.4, 4)$, \ding{55}     & 28.21 & 0.82 \\
\bottomrule
\end{tabular}
}
\hfill
\adjustbox{width=0.49\linewidth}{
\centering
\begin{tabular}{lcc}
\toprule
\textbf{blur, noise} & \textbf{PSNR}$\uparrow$ & \textbf{SSIM}$\uparrow$ \\
\midrule
$(0.4, 1, 3)$, \ding{55}          & 28.78 & 0.84 \\
$(0.4, 1, 3)$, \ding{51}   & 27.30 & 0.79 \\
\bottomrule
\end{tabular}
} \vspace{-1mm}
\caption{Evaluation of our synthesis under various imaging qualities.}\vspace{-2mm}
\label{tab:blur_noise_synthesis}
\end{table}

\paragraph{Baselines.}
For tagged-to-cine synthesis, we compare: (i)~\emph{LowpassFuse}, which applies a 1D Gaussian low-pass filter along each tag orientation to remove tags and averages the results; and (ii)~\emph{HARP Demodulation}~\cite{osman2000imaging,PVIRA}, which averages the magnitudes of HARP-filtered complex images, using a band-pass filter centered at the tag frequency with a radius equal to half the true tagging frequency.
For motion estimation, we compare both learning-based---LKUnet~\cite{LKUnet2022} and DeepTag~\cite{ye2021deeptag,ye2023sequencemorph}---and optimization-based methods, such as SyN~\cite{AvantsGee2008} and DRIMET~\cite{bian2024drimet}.
DeepTag (originally 2D, grid-tagged) is adapted to 3D by replacing 2D with 3D convolutions and forming grid tags via multiplicative composition of three orthogonal line-tag images. 
For fair testing, learning-based methods are pretrained on tagged sequences generated from 150 T1w brain images from HCP~\cite{van2013wu} data. 
DRIMET registers sinusoidal-transformed HARP~(sHARP) images, and we use SyN as its registration backend. We rely on author-provided implementations for HARP Demodulation\footnote{\scriptsize \url{iacl.ece.jhu.edu/index.php?title=HARP_FAQ}}, LKUnet\footnote{\scriptsize\url{github.com/xi-jia/LKU-Net}}, DeepTag\footnote{\scriptsize\url{github.com/DeepTag/cardiac_tagging_motion_estimation}},  DRIMET\footnote{\scriptsize\url{github.com/jasonbian97/DRIMET-tagged-MRI}}, and SyN~\cite{ANTs}.
Smoothness hyperparameters are selected via grid search on a held-out validation set of 20 tagged sequences.

\subsection{Tag-to-Cine Results}
We first assess how well our method reconstructs high-resolution anatomy while maintaining temporal consistency.
We compare against LowpassFuse and the HARP-based demodulation baseline, reporting SSIM and PSNR at two time points in Table~\ref{tab:synthesis_metrics}.
Our method yields substantially better synthesis quality, including at later frames where tag fading and deformation become pronounced (e.g., $t{=}6$), as shown in Fig.~\ref{fig:cine_compare}.
Unlike both baselines, which require the tag frequency to be known a priori to design band-pass filters (e.g., center and radius of the HARP filter), our method does \emph{not} assume this information and instead estimates it directly.
LowpassFuse produces visible aliasing artifacts (blocky streaks) due to the difficulty of manually selecting the cutoff frequency: choosing it too low oversmooths the image, while choosing it too high leaves residual tag patterns that alias.
Similar issues affect the HARP demodulation approach. As seen in the Fourier spectra of the inputs (Fig.~\ref{fig:cine_compare}, Col.~2), deformation progressively smears the harmonic peaks, making them harder to isolate with any fixed band-pass filter.

\paragraph{Test under various imaging qualities.}
Table~\ref{tab:blur_noise_synthesis} and Fig.~\ref{fig:ours_various_cond} (Appendix) show that our method remains robust across varying blur kernels and noise levels. 

\begin{figure}
    \centering
    \includegraphics[width=1\linewidth]{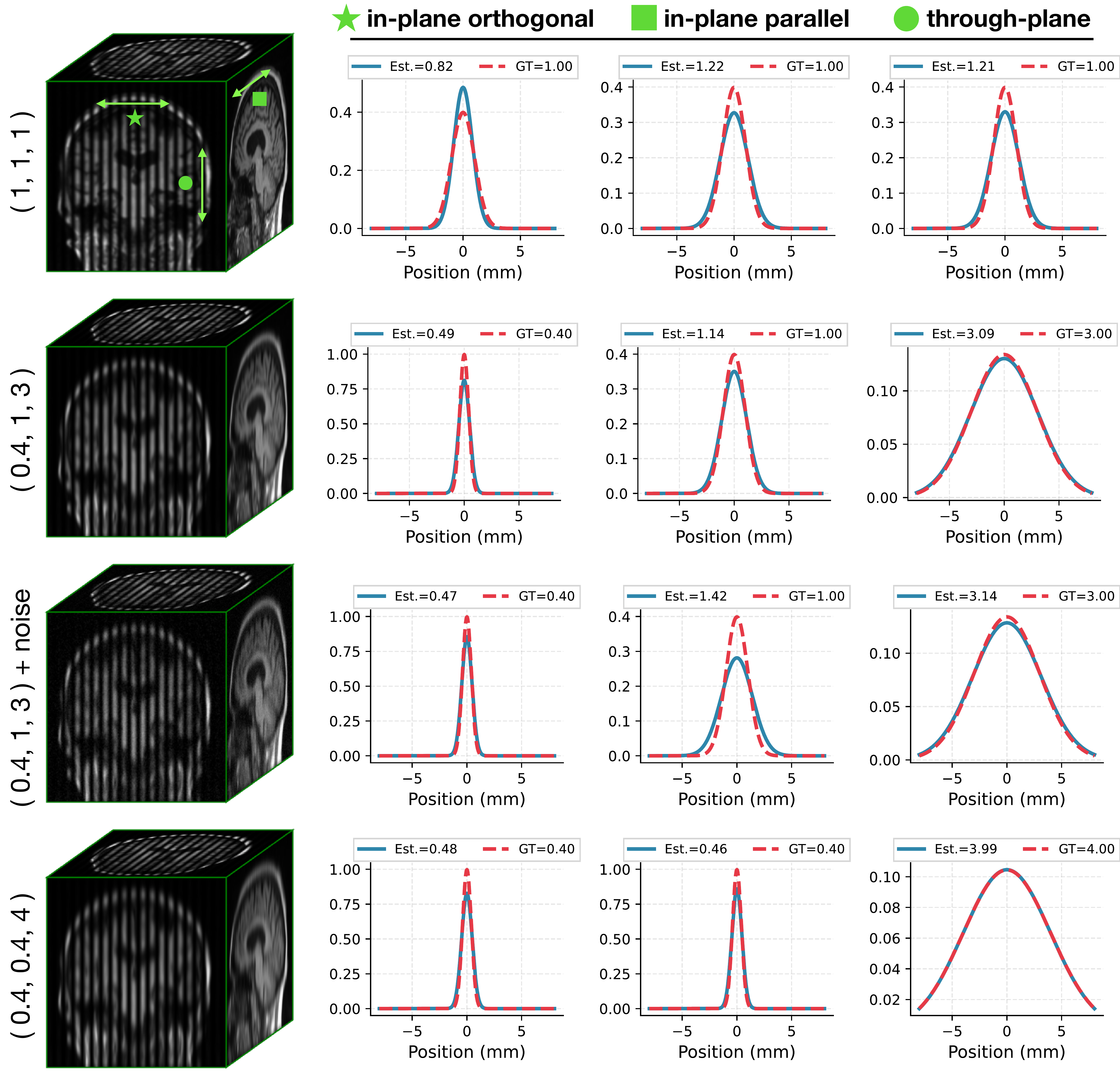}
    \caption{Results of estimated point spread functions (PSFs) under different blurring settings 
and noise levels.}
    \label{fig:PSF_est}
\end{figure}

\paragraph{Estimate PSF.} 
Instead of learning a data-driven prior over PSFs~\cite{bora2018ambientgan,levac2025double}---which often suffers from domain shift---we blindly and jointly estimate imaging PSFs during optimization. 
Fig.~\ref{fig:PSF_est} shows PSF reconstructions across different blur and noise conditions.

\subsection{Motion Estimation Results}
\begin{figure}
    \centering
    \includegraphics[width=1.0\linewidth]{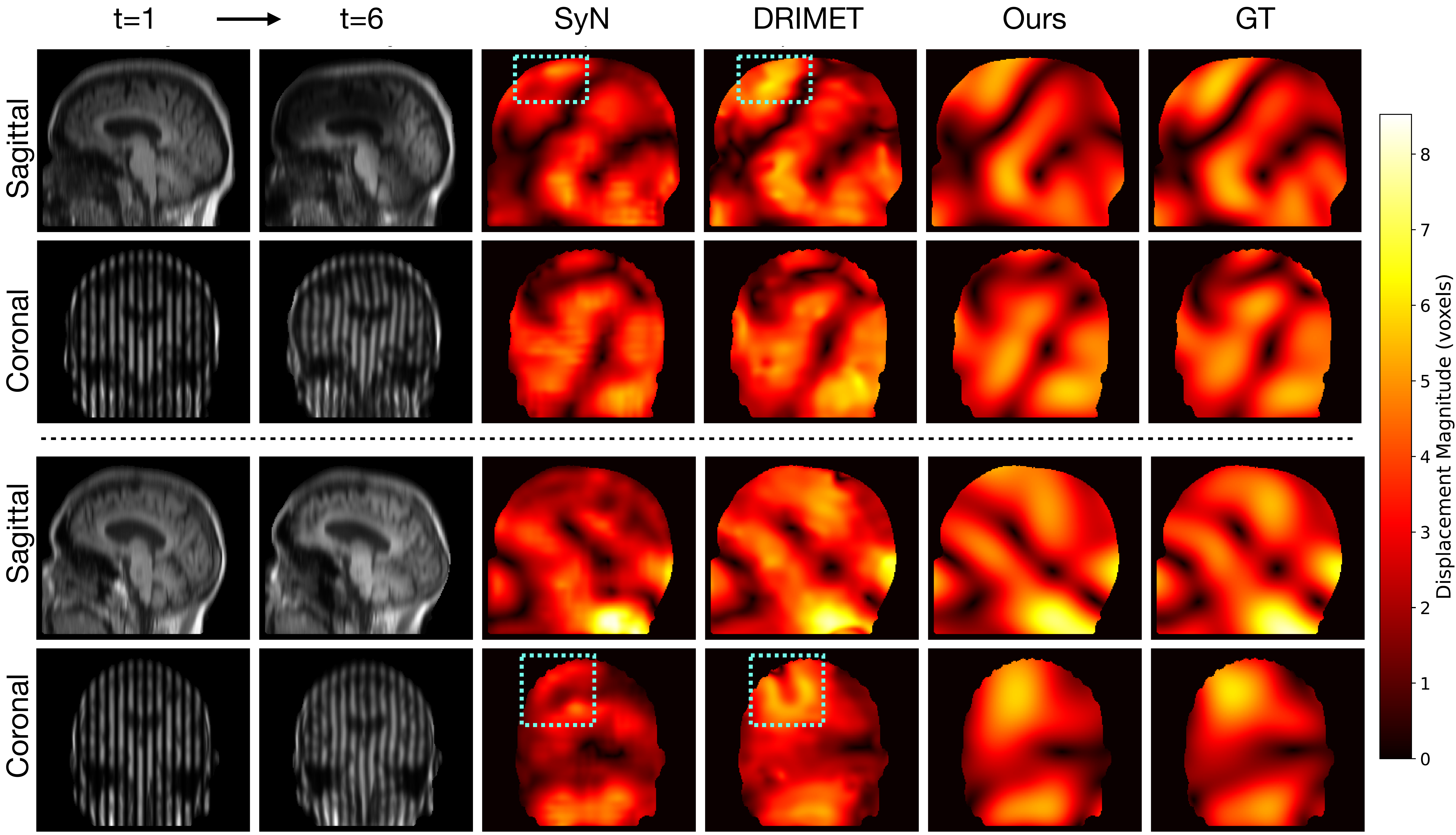}
    \caption{Motion magnitude comparison across methods. Cols.~1-2: input tagged MRI at two time points. Last column: ground-truth motion magnitude. Both sagittal and coronal views are shown.}
    \label{fig:motion_quali}
\end{figure}

Given a sequence of tagged MRI, we evaluate the resulting motion accuracy using the mean EPE, EPE@95, and NegDet. As shown in Table~\ref{tab:epe_comparison}, learning-based registration (LKUnet, DeepTag), which directly take raw tagged MRI as input, suffer from local contrast changes due to tag fading, and domain shift when applied to new datasets.
Among optimization-based baselines, SyN improves both average and tail errors over learning-based methods, and DRIMET further reduces EPE by explicitly leveraging sHARP representations. Our method achieves the lowest EPE, while also the lowest EPE@95. This indicates fewer large local errors, which is crucial for reliable strain estimation for biomechanical studies.
All optimization-based methods report NegDet $<0.001\%$, confirming that they produce nearly diffeomorphic fields.  Qualitative comparisons in Fig.~\ref{fig:motion_quali} show that our estimated motion magnitudes follow the ground truth more closely in large deforming regions and along complex cortical folds, whereas competing methods tend to underestimate peak motion or introduce spurious spatial fluctuations.

\subsection{Validation on Real Tagging Data}

\begin{figure}[t]
    \centering
    \includegraphics[width=1.0\linewidth]{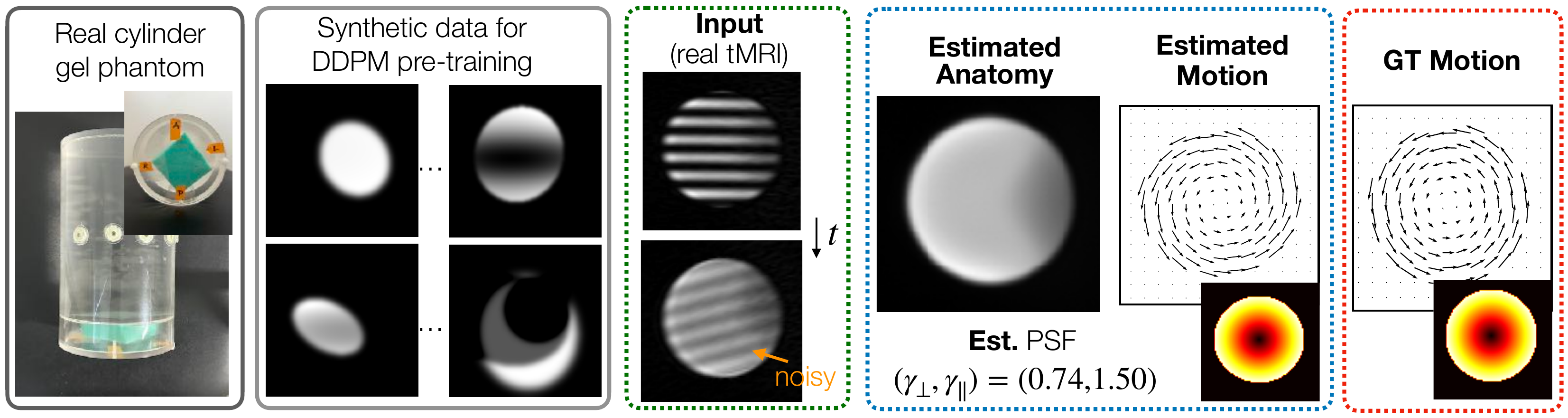}
    \vspace{-1.8em}
    \caption{\textbf{Real tagged MRI: gel cylinder.}}
    \label{fig:phantom-exp}
    \vspace{-0.7em}
\end{figure}

To assess generalization beyond simulation, we evaluate InvTag on real tagging acquisitions where real imaging artifacts are present. We tested InvTag on real tagged MRI of a rotating \textbf{gel phantom} (Fig.~\ref{fig:phantom-exp}). Notably, the diffusion prior was pre-trained solely on synthetic ellipses. Despite this train-test domain gap, InvTag successfully recovered the anatomy and estimated the rotational motion under real scanner acquisition effects: field inhomogeneity, blurring, non-Gaussian noise, and severe tag fading.

\subsection{Ablation studies}
We evaluate the contribution of three components: PSF estimation, tag-fading estimation, and the coordinate descent diffusion prior (CDDP). Table~\ref{tab:ablation-components} summarizes the results. (A)~Removing PSF estimation noticeably degrades synthesis quality, showing that explicitly estimating the PSF is important to bridge the resolution gap between low-resolution tagged MRI and the high-resolution diffusion prior. (B)~Disabling fading estimation, i.e., keeping a time-invariant tag pattern, yields worse tracking, indicating that modeling time-varying tag fading is necessary to disentangle contrast changes from tissue motion. (C)~Replacing CDDP with simultaneous optimization of all parameters in~\eqref{eq:ml-nuisance} using Adam optimizer leads to failure due to the ill-posedness of the nonlinear blind inverse problem. The full model (``Ours'') with all three components enabled achieves the best across all metrics.

\begin{table}
\scriptsize
\centering
\begin{tabular}{lccccc}
\toprule
\textbf{Method} & \textbf{Opt.} & \textbf{EPE} $\downarrow$ & \textbf{EPE@95} $\downarrow$ & \textbf{NegDet(\%)} $\downarrow$ \\
\midrule
LKUnet~\cite{LKUnet2022} & \ding{55} & 1.35 $\pm$ 0.65 & 2.94 $\pm$ 1.30 & 0.043 \\
DeepTag~\cite{ye2021deeptag} & \ding{55} & 1.27 $\pm$ 0.58 & 2.97 $\pm$ 1.13 & 0.060 \\ 
\cdashline{1-5}\noalign{\vskip 0.6ex}
SyN~\cite{AvantsGee2008}     & \ding{51} & 1.06 $\pm$ 0.31 & 2.41 $\pm$ 0.80 & $<0.001$ \\
DRIMET~\cite{bian2024drimet}  & \ding{51} & 0.79 $\pm$ 0.14 & 1.61 $\pm$ 0.53 & $<0.001$ \\ 
\cdashline{1-5}\noalign{\vskip 0.6ex}
Ours    & \ding{51} & \textbf{0.60} $\pm$ \textbf{0.11} & \textbf{1.31} $\pm$ \textbf{0.22} & $<0.001$ \\
\bottomrule
\end{tabular} \vspace{-1mm}
\caption{Quantitative comparison of motion estimation accuracy. 
“Opt.” denotes whether the method is optimization-based.}
\label{tab:epe_comparison}
\end{table}

\begin{table}[t]
\centering
\scriptsize
\adjustbox{width=0.99\linewidth}{
\begin{tabular}{rccccccc}
\toprule
\textbf{\#} &
\textbf{PSF} &
\textbf{fading} &
\textbf{CDDP} &
\textbf{PSNR}$\uparrow$ &
\textbf{SSIM}$\uparrow$ &
\textbf{EPE}$\downarrow$ &
\textbf{EPE@95}$\downarrow$ \\
\midrule
A     & \ding{55} & \ding{51} & \ding{51} & 27.27 & 0.69 & 0.62 & 1.41 \\
B     & \ding{51} & \ding{55} & \ding{51} & 28.21 & 0.80 & 0.71 & 1.56 \\
C     & \ding{51} & \ding{51} & \ding{55} & 22.05 & 0.46 & 1.57 & 2.73 \\
Ours  & \ding{51} & \ding{51} & \ding{51} & \textbf{28.40} & \textbf{0.83} & \textbf{0.60} & \textbf{1.31} \\
\bottomrule
\end{tabular}
} \vspace{-1mm}
\caption{Ablation study of method components. Std is omitted.}
\label{tab:ablation-components}
\end{table}

\paragraph{Influence of diffusion prior strength.}
Sweeping the prior weight $\rho^*$ in~\eqref{eq:dps-functional} shows that InvTag is robust across a wide operating range ($\rho^* \in [50, 200]$); only an overly weak data term ($\rho^*{=}1$) causes hallucinated reconstructions. See Appendix~\ref{sec:prior-strength} for the full sweep.

\paragraph{Empirical identifiability and convergence.}
Across $N{=}5$ runs with random initializations, estimated PSF and tag parameters exhibit negligible variance and motion accuracy remains stable, confirming reliable convergence of CDDP. Full results are in Appendix~\ref{sec:convergence}.

\section{Discussion}
\paragraph{Limitations \& future work.}
InvTag currently takes a long time to run due to repeated diffusion sampling and the optimization of PINN networks. Acceleration techniques for diffusion model sampling and multi-GPU parallelization could substantially reduce runtime.
InvTag currently assumes sinusoidal tags; generalizing to higher-order and grid tags would broaden applicability.
Extending evaluation to real scans such as cardiac MR tagging is important future work. 
Finally, incorporating uncertainty quantification into the inversion could improve interpretability and clinical trust.

\paragraph{Conclusion.}
We presented \emph{InvTag}, a \emph{nonlinear} and \emph{blind} inverse framework for 3D tagged MRI that jointly recovers high-resolution anatomy, synthesizes cine images, and estimates diffeomorphic motion. InvTag is a \emph{unified} formulation, addressing challenges---tag-to-cine synthesis, super-resolution, and motion estimation---that have long been treated separately. The results on tagged head MRI demonstrate the synergy of jointly modeling anatomy, imaging parameters, and motion. We also showcase that generative priors can be used for solving \emph{nonlinear} \emph{blind} inverse problems in medical imaging.

\paragraph{Acknowledgements.}
This work was supported in part by National
Institutes of Health (NIH) grant U01NS112120 and Johns Hopkins Discovery Award.

{
    \small
    \bibliographystyle{ieeenat_fullname}
    \bibliography{main}
}

\clearpage

\clearpage
\appendix

\twocolumn[{
\begin{center}
\includegraphics[width=\linewidth]{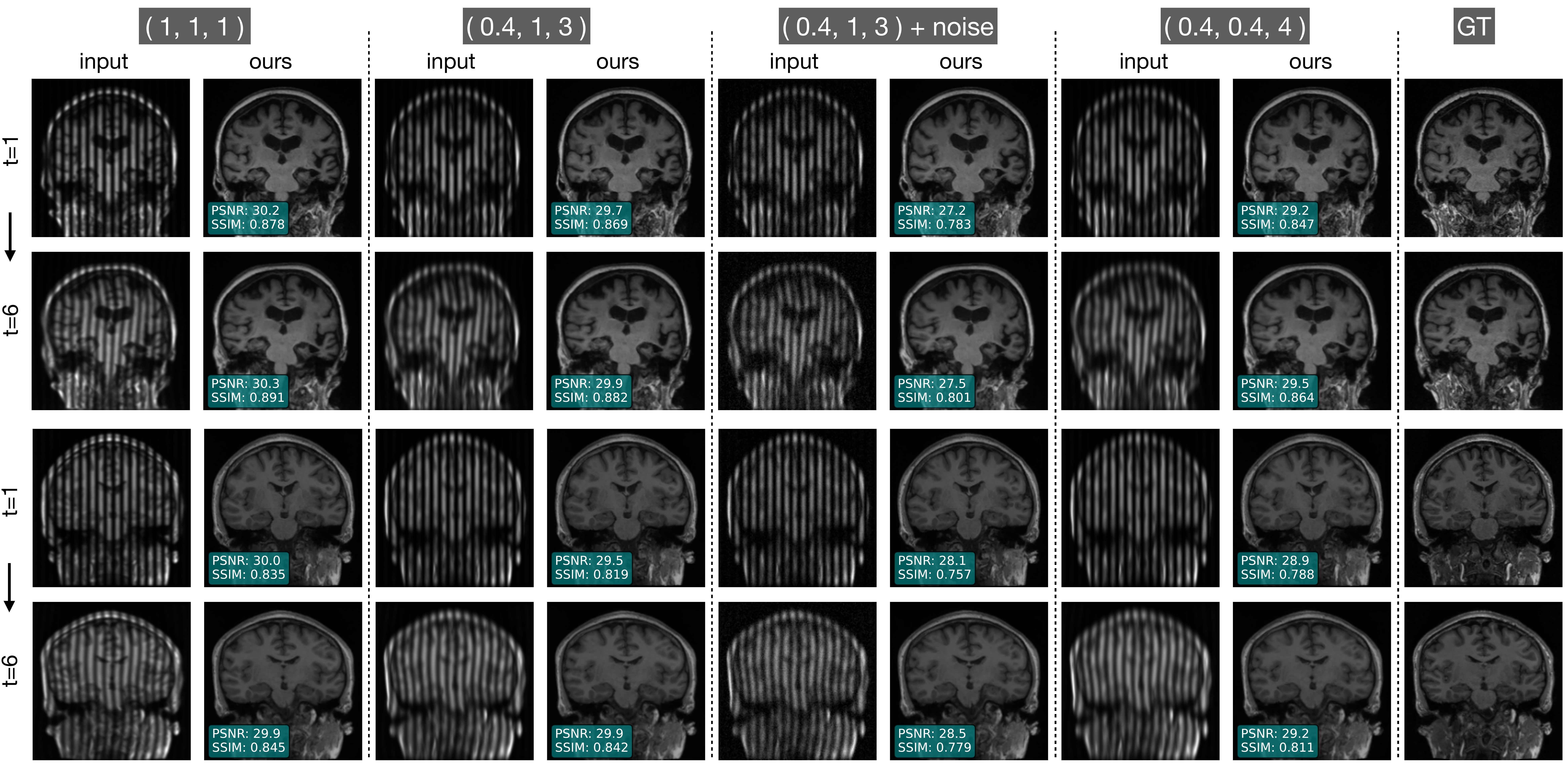}
\end{center}
\vspace{-0.5cm}
\captionsetup{type=figure}
\captionof{figure}{Qualitative results of estimated reference anatomy under four blurring settings
\((\gamma_\perp, \gamma_\parallel, \gamma_{\mathrm{thru}})\) and two noise levels: no noise and $\mathcal{N}(0,0.01)$.
Only one input tagged image is shown for brevity.
The ground-truth anatomy is shown in the last column.}
\label{fig:ours_various_cond}
\vspace{0.3cm}
}]

\begin{table}[t]
\centering
\scriptsize
\resizebox{0.95\linewidth}{!}{%
\begin{tabular}{lc|lc|lc}
\toprule
\multicolumn{2}{c|}{\textbf{PSF Params}} & \multicolumn{2}{c|}{\textbf{Tag Params}} & \multicolumn{2}{c}{\textbf{Metrics}} \\
\midrule
$\gamma_\perp$ & $0.42 \pm 0.03$ & $\alpha_1$ & $0.74 \pm 0.09$ & PSNR & $28.14 \pm 0.11$ \\
$\gamma_\parallel$ & $1.03 \pm 0.06$ & $\alpha_2$ & $10.1 \pm 0.00$ & SSIM & $0.81 \pm 0.02$ \\
$\gamma_{\text{thru}}$ & $2.95 \pm 0.12$ & $\alpha_3$ & $-1.14 \pm 0.00$ & EPE & $0.63 \pm \mathbf{0.03}$ \\
 & & $\alpha_4$ & $0.22 \pm 0.07$ & & \\
\bottomrule
\end{tabular}%
}
\caption{\textbf{Empirical identifiability analysis.} Mean $\pm$ standard deviation across $N{=}5$ independent runs with random initializations.}
\label{tab:stability}
\end{table}

\section{Implementation Details}
\label{sec:impl-details}

Volumes are center-cropped and/or padded to $192 \times 224 \times 192$ voxels and normalized to $[0, 1]$ using the min–max of the 1$^{\text{\scriptsize{st}}}$ time frame.
The forward operator implements the orientation-specific PSF as an anisotropic 3D Gaussian via separable 1D convolutions. The spatial warp is a differentiable spatial transformer~\cite{jaderberg2015spatial} (\texttt{torch.grid\_sample}).
The DDPM is a five-level 3D U-Net (encoder–decoder with skip connections).
Channel widths per level are $(16,32,64,128,256)$, with two residual blocks per level.
$\rho$ in~\eqref{eq:dps-functional} is set to $100/\mathcal{L}_\mathrm{rec}^t$.
The PINN $\theta_t$ is a fully connected network with three hidden layers (128 neurons each) that outputs a stationary velocity field; diffeomorphic maps are obtained by 7 scaling-and-squaring steps.
Adam uses $(\beta_1,\beta_2)=(0.9,0.999)$ with a fixed learning rate of $5e^{-4}$, and differential evolution (from \texttt{scipy}) uses a population of 30, strategy \texttt{best1bin}, relative tolerance $10^{-2}$, and a maximum of 200 iterations per block.
Motion parameters $\theta_t$ are trained for a fixed 2000 Adam steps per inner loop. Differential evolution terminates when the relative change in the objective falls below $10^{-2}$.

\section{Qualitative Results Under Various Imaging Conditions}
\label{sec:various-cond}
Figure~\ref{fig:ours_various_cond} shows qualitative results of estimated reference anatomy under four blur configurations and two noise levels. Across all settings---including strong through-plane blur $(0.4,0.4,4)$ and additive Gaussian noise---the recovered anatomy preserves cortical detail and closely matches the ground truth.

\section{Influence of Diffusion Prior Strength}
\label{sec:prior-strength}
\begin{figure}[t]
    \centering
    \includegraphics[width=1\linewidth]{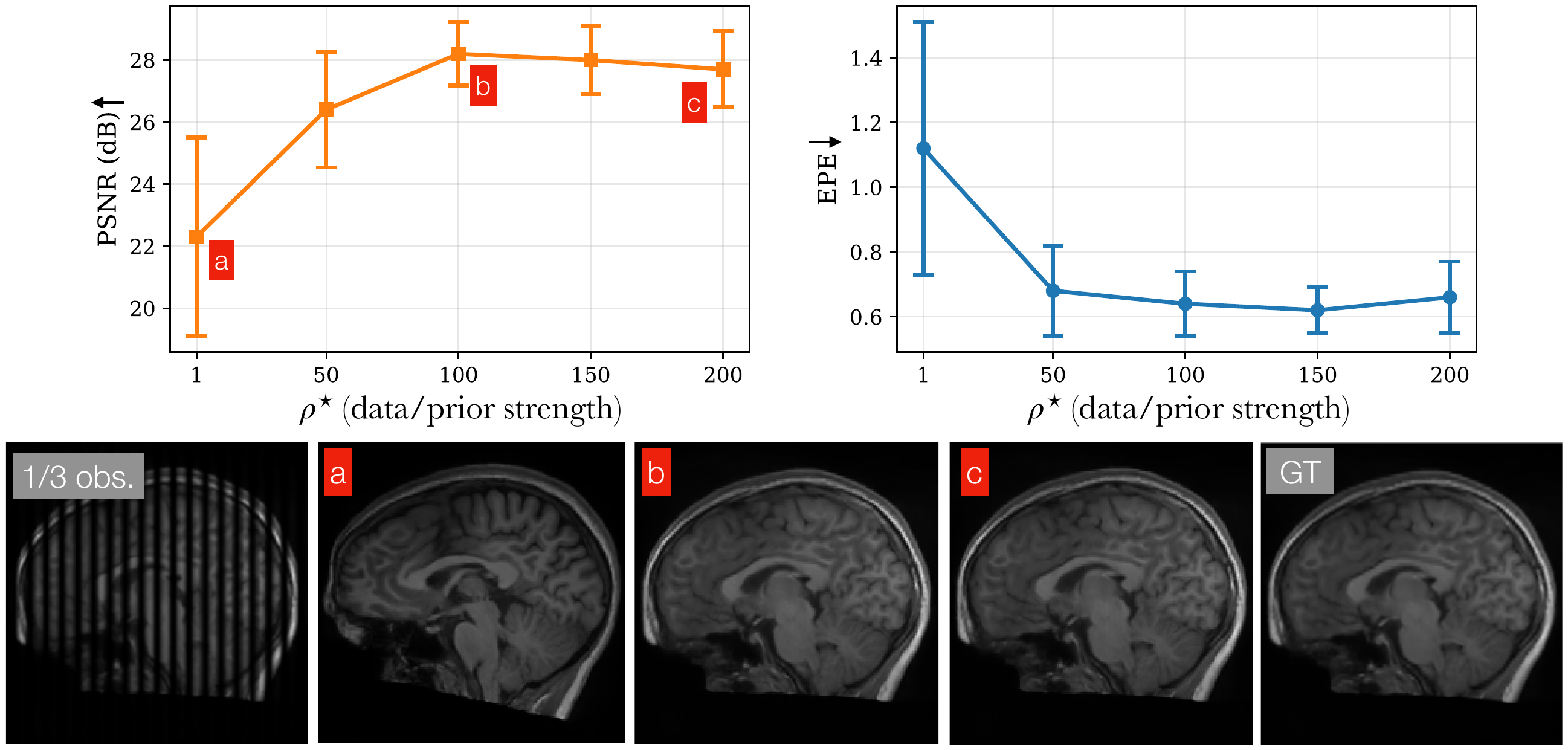}
    \caption{\textbf{Data fidelity vs.\ prior strength.} A sweep from strong prior ($\rho^*=1$) to strong data fidelity ($\rho^*=200$).}
    \label{fig:prior-strength}
\end{figure}

We analyzed the trade-off between data fidelity and diffusion prior strength by sweeping the weight $\rho:=\rho^* / \mathcal{L}_\mathrm{rec}^t$ in~\eqref{eq:dps-functional}. Fig.~\ref{fig:prior-strength} shows that the method is robust across a wide operating range ($\rho^* \in [50, 200]$). An overly weak data term ($\rho^*=1$) causes the reconstruction to collapse onto the prior manifold, producing hallucinated structures (Fig.~\ref{fig:prior-strength}a), whereas appropriate weighting effectively balances measurement consistency with anatomical plausibility.

\section{Empirical Identifiability and Convergence}
\label{sec:convergence}
While theoretical convergence proofs for this nonlinear blind inversion are challenging, we provide empirical evidence of stability. Tab.~\ref{tab:stability} shows that across $N{=}5$ runs with random initializations, estimated parameters exhibit negligible variance and motion estimation remains stable (EPE$\,{=}\,0.63 \pm 0.03$) despite the problem's ill-posedness and non-uniqueness. This complements the ablation in Tab.~\ref{tab:ablation-components}C, which shows that joint optimization (without CDDP) fails, whereas our coordinate descent scheme converges reliably.

\end{document}